%% file: nrecc.tex
\documentclass[prd,nofootinbib,twocolumn]{revtex4}
\usepackage{amssymb} \usepackage{graphicx} \input{defs}
\begin{document}

\title{Circularization and Final Spin in Eccentric Binary Black Hole
Inspirals}

\author{Ian Hinder} \author{Birjoo Vaishnav} \author{Frank Herrmann}
\author{Deirdre M. Shoemaker}
\author{Pablo Laguna}
\affiliation{Center for Gravitational Wave Physics, The Pennsylvania
  State University, University Park, PA 16802}

\begin{abstract} We present results from numerical relativity
  simulations of equal mass, non-spinning binary black hole inspirals
  and mergers with initial eccentricities $e\le 0.8$ and coordinate
  separations $D \ge 12\,M$ of up to 9 orbits (18 gravitational wave
  cycles). We extract the mass $M_\mathrm{f}$ and spin $a_\mathrm{f}$
  of the final black hole and find, for eccentricities $e\lesssim
  0.4$, that $a_\mathrm{f}/M_\mathrm{f} \approx 0.69$ and
  $M_\mathrm{f}/M_\mathrm{adm} \approx 0.96$ are {\em independent} of
  the initial eccentricity, suggesting that the binary has
  circularized by the merger time. For $e \gtsim 0.5$, the black holes
  plunge rather than orbit, and we obtain a maximum spin parameter
  $a_\mathrm{f}/M_\mathrm{f} \approx 0.72$ around $e = 0.5$. 
\end{abstract}

\maketitle

The field of \nr{} has now entered a stage where \bbh{} simulations
can reliably be used to investigate a vast range of interesting
phenomena. Studies have produced gravitational waveforms from binary
systems in essentially circular
orbits~\cite{Baker:2006yw,2007gr.qc.....1164C,Hannam:2007ik,boyle-2007},
involving spinning black holes as well as unequal mass systems.  The
level of numerical accuracy achieved by these codes is
impressive~\cite{husa-2007,boyle-2007}, and in some of these studies,
the initial binary separations were such that it was feasible to
directly compare with \pnw{}
waveforms~\cite{Hannam:2007ik,boyle-2007}. Other examples of exciting
new results in \nr{} are investigations of the kick imparted to the
final
\bh{}~\cite{Gonzalez:2006md,2007ApJ...661..430H,Baker:2006vn,2007PhRvL..99d1102K,2007gr.qc.....1164C,Gonzalez:2007hi,Campanelli:2007cga},
the spin dynamics of the merging
\bh{s}~\cite{Campanelli:2006uy,2007arXiv0706.2541H}, and, of relevance
to the present work, the merger threshold between bound and unbound
\bbh{}s~\cite{pretorius-2006-23,pretorius-2007}, and studies of the
result of a circular inspiral of spinning
\bh{}s~\cite{2007arXiv0707.2559P,2007arXiv0708.3999R,2007arXiv0710.3345R}.
All of this has been possible since the pioneering work
of~\citet{Bruegmann:2003aw,2005PhRvL..95l1101P,Campanelli:2005dd,Baker:2005vv}.

It is well known that gravitational radiation leads to circularization
of a binary system~\cite{Peters:1964}. In this work, we study this
circularization in the nonlinear regime.  In Ref.~\cite{Baker:2006yw},
it was found that, for equal-mass, non-spinning \bh{s} initially in
quasi-circular orbits, the merger produced a \bh{} with spin parameter
$a_\mathrm{f}/M_\mathrm{f} \sim 0.69$, which, within the accuracy of
the results, was independent of the initial separation.  

In this \emph{Letter}, our main goals are (1) to investigate whether
sufficient eccentricity is lost during the late stages of inspiral to
circularize the orbit and exhibit the same \emph{universality} as in
the circular case and (2) to extract the spin parameter and mass of
the final \bh{}, and compare the values with those from circular
inspirals.

\begin{figure}
\includegraphics[scale=1.0]{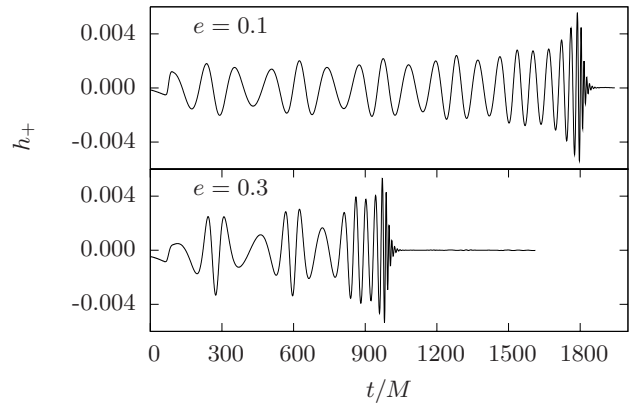}
\caption{Waveform polarizations $h_+$ for the cases $e=0.1$ and
  $e=0.3$}
\label{fig:strain}
\end{figure}

Although isolated stellar mass \bbh{s} will have completely
circularized by the time they are observable by ground-based
interferometers, scenarios have been suggested for which \bbh{s} in
eccentric orbits are not only astrophysically interesting but also
could be detected by space- or ground-based
interferometers~\cite{2002ApJ...576..894M,2003ApJ...598..419W}.  For
instance, galactic mergers leave behind massive \bbh{s} that likely
interact with a gaseous environment.  A gaseous-gravitational driven
inspiral could yield a \bbh{} arriving at the last few orbits and
merger with a non-vanishing eccentricity.  An observation of the
gravitational waves from an eccentric \bbh{} merger will allow us to
determine the amount of angular momentum lost to gas
and, in particular, the gravitational torques between
the binary and a circumbinary disc that affect the eccentricity of the
binary~\cite{2005ApJ...634..921A}.


{\em Methods:} We construct initial data using the puncture
approach~\cite{Anninos:1995vf}, which requires specifying the
coordinate locations and momenta for the two \bh{s}.  For ``circular''
orbits, we follow~\cite{Husa:2007rh}.
For eccentric models, we use the conservative 3PN expressions in
Ref.~\cite{2006PhRvD..73l4012K}.  These expressions require the
specification of the eccentricity $e$ and the mean motion $n =
2\pi/P_r$, where $P_r$ is the radial (pericenter to pericenter)
orbital period.  There are three PN eccentricities, which are the same
to 1 PN order, and we choose $e_t$, which appears in the PN Kepler
equation, following Ref.~\cite{2006PhRvD..73l4012K}.  It is important
to keep in mind that the eccentricities we quote (and we use them also
to label the models) are to be taken only as a guide to the
eccentricity in the initial data, as the \pnw{} expressions used do
not include radiation reaction, and the \pnw{} parameters are in a
different coordinate system to the puncture initial data.

We construct a family of initial data by fixing $n = 0.01625/M$ ($P_r
\sim 387 M$) and varying $e$ in the range $0.05-0.8$ (note that to 2PN
order, this means that the systems have the same binding energy and
that, at high eccentricities, there are portions of the orbit for
which the \pnw{} condition $v/c \ll 1$ is no longer valid). The binary
separation $D$ is determined from Eq.~(23) in
Ref.~\cite{2006PhRvD..73l4012K}, and the tangential linear momentum,
$P/M$, of each \bh{} at apocenter is obtained from $J = P D$, where
$J$ is the total angular momentum computed as a \pnw{} expansion in
$n$ and $e$ (Eq.~(21) in Ref.~\cite{2006PhRvD..73l4012K}). The bare \bh{}
masses $m_{1,2}$ are chosen to make the irreducible \bh{} masses
$M_{1,2} = 0.5$ (i.e. $M=M_1+M_2 = 1$).  Table~\ref{tbl:ID} provides
the initial data parameters.

\begin{table}
  \begin{center}
\begin{tabular}{c|cc||c|cc}
\hline
\hline
$e$ & $D/M$ & $P_{1,2}/M$ & $e$ & $D/M$ & $P_{1,2}/M$ \\
\hline 
0.00 &  12.000 & 0.0850  & 0.40 &  18.459 & 0.0498 \\
0.05 &  12.832 & 0.0792  & 0.50 &  20.023 & 0.0429 \\
0.10 &  13.645 & 0.0741  & 0.60 &  21.539 & 0.0361 \\
0.15 &  14.456 & 0.0695  & 0.70 &  22.955 & 0.0292 \\
0.20 &  15.264 & 0.0651  & 0.80 &  24.072 & 0.0214 \\
0.30 &  16.870 & 0.0571  &--    &--       & -- \\
\hline
\hline
\end{tabular}
  \end{center}
  \caption{\emph{Initial data parameters:} The runs are labeled by
    their initial eccentricity $e$.  The \bh{s} have linear momenta
    $\pm P_{1,2}/M$ and are separated by a coordinate distance $D/M$.  }
\label{tbl:ID}
\end{table}

The numerical simulations and results in this work were obtained with
the same infrastructure
used in our previous \bbh{} studies (see Ref.~\cite{vaishnav-2007}
for full details).  We have evolved the circular model at three
different resolutions (finest grid spacings of $M/38.7,\, M/51.6$ and
$M/64.5$).  We obtain approximately fourth order convergence in the
total energy and angular momentum radiated, consistent with the
designed 4th order accuracy.

\begin{figure}
\includegraphics[scale=1.0]{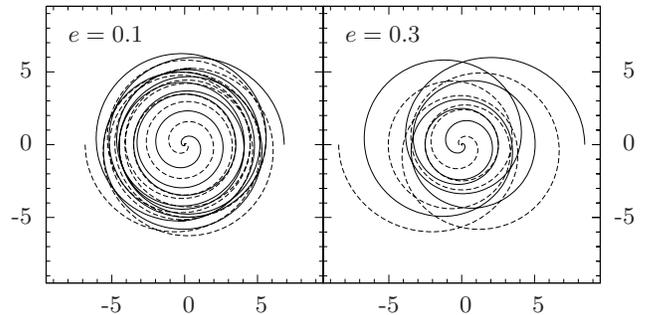}
\caption{Inspiral tracks for initial eccentricity $e=0.1$
  (left panel) and $e=0.3$ (right panel).}
\label{fig:tracks}
\end{figure}

{\em Results:} In Figs.~\ref{fig:strain} and \ref{fig:tracks} we
display the gravitational wave strains and coordinate inspiral tracks
for $e=0.1$ and $e=0.3$. It is evident that the difference in initial
eccentricity has a large effect during the inspiral. Qualitatively,
the case with larger eccentricity exhibits a more rapid
inspiral\cite{Peters:1964}. However, at some point both systems enter
a ``circular'' plunge, hinting that circularization may have occurred.
We find that the simulations with $e \ge 0.5$ show plunge-type rather
than orbital-type behavior in the coordinate motion from the very
start.  Note that the tracks shown in Fig.~\ref{fig:tracks} represent
the {\em coordinate} positions of the individual \bh{}s, and once a
common horizon forms, they are less meaningful.

\begin{figure}[b]
\includegraphics[scale=1.0]{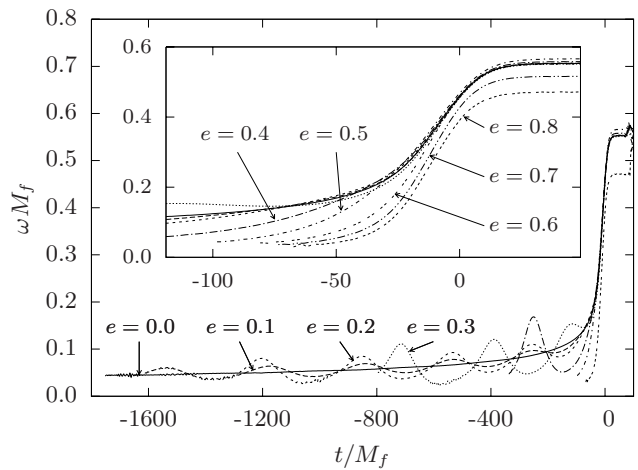}
\caption{Frequency of the $\ell=2,\,m=2$ mode of the \nps{}
  radiation scalar $r\,\Psi_4$}
\label{fig:freq}
\end{figure}

We now consider the emitted radiation and focus on the dominant
$\ell=2,\,m=2$ mode of the complex \nps{} quantity $\Psi_4 =
A(t)\exp{(-i\varphi(t))}$. To compare the orbits, we apply a time
shift to $A$ and $\varphi$, so that the maximum of $A$ (i.e.~the peak
of the amplitude of the gravitational wave) is at $t/M_\mathrm{f} = 0$
in each simulation. In Fig.~\ref{fig:freq}, we plot the shifted
amplitudes and frequencies $\omega = d\varphi/dt$ extracted at $r =
70\,M$. The cases displayed are those with eccentricities $e=0-0.5$ in
steps of $0.1$ and $e=0.8$.

In Fig.~\ref{fig:freq}, the oscillations and growth in $\omega$ at
early times (inspiral) can be in general terms understood from simple
Newtonian considerations.  That is, ignoring radiation reaction, the
oscillations (i.e.~amplitude and period) in $\omega$ are a direct
consequence of the eccentricity and not present in the $e=0$ case.
The period of these oscillations is the period $P_r$ of the extrema in
the separation, and the amplitude of the oscillations increases with
$e$.  The addition of radiation reaction leads to an overall growth of
$\omega$ with time due to the energy and angular momentum loss, and
this is clearly visible in the figure.  The amplitude of the
oscillations in $\omega$ should decrease with time, corresponding to a
reduction in eccentricity.  However, over these timescales, it is
difficult to separate this effect from the secular increase in
$\omega$.  Also consistent with the predictions in~\cite{Peters:1964},
the higher eccentricity evolutions merge more quickly.  We note that
in the Newtonian case, we would observe that $P_r = P_\phi$, where
$P_\phi$ is the time the binary takes to complete one revolution in
the angular coordinate $\phi$.  Due to the effects of precession
caused by general relativity, the two periods are very different (this
can be seen from \pnw{} equations).

During the merger or plunge phase, $\omega$ increases dramatically and
then levels off, signaling that the \bbh{} has merged.  After this
point, $\omega$ remains constant, a direct consequence of the \qnm{}
ringing of the final \bh{.} 

As mentioned before, one of the objectives of this work is to
investigate whether a given initial data configuration will
circularize before it merges.  By this we mean that the radiation from
the late stages of the evolution is identical to that from an orbit
which started with zero eccentricity.  We see in Fig.~\ref{fig:freq}
that the low $e$ evolutions approach the same final state as a
circular orbit at the very late stage of inspiral.  The $\omega$s from
different eccentricities near $t/M_\mathrm{f} = 0$ seem to be
indistinguishable for low enough eccentricity. In order to investigate
this in more detail, in the inset of Fig.~\ref{fig:freq} we focus on
the plunge stage.  Here we plot eccentricities $e=0-0.8$ in steps of
$0.1$. Up until $e=0.4$ and after $t/M_\mathrm{f} \approx -50$, the
frequencies $\omega$ from each run follow each other.  Noticeable
differences start showing for $e\ge 0.5$, which is the first
configuration to plunge immediately without orbiting first.

\begin{figure}
\includegraphics[scale=1.0]{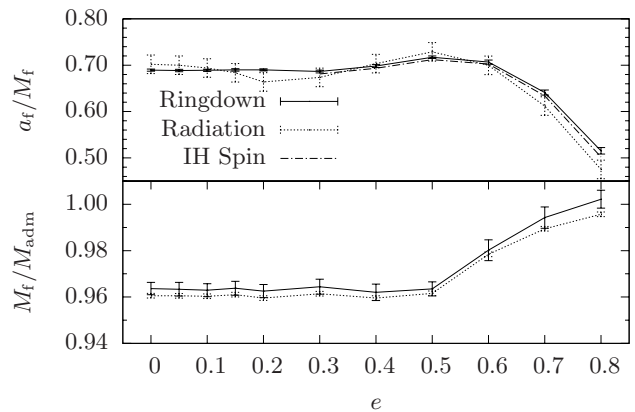}
\caption{Plots of $M_\mathrm{f}/M_\mathrm{adm}$ and
  $a_\mathrm{f}/M_\mathrm{f}$ as functions of the initial eccentricity
  $e$.  Note that the error bars shown here on the radiation
  quantities for eccentric runs are taken from the low resolution
  circular case and thus should be treated as indicative only.}
\label{fig:massspin}
\end{figure}

We now discuss $M_\mathrm{f}$ and $a_\mathrm{f}$, computed using two
independent methods. In one method, they are obtained from the
radiated energy and angular momentum using $M_\mathrm{f} =
M_\mathrm{adm} - E_\mathrm{rad}$ and $a_\mathrm{f}/M_\mathrm{f} =
(J_\mathrm{adm} - J_\mathrm{rad})/M^2_\mathrm{f}$.  In the second
method, $M_\mathrm{f}$ and $a_\mathrm{f}$ are computed from the \qnm{}
frequencies \cite{BertiLISA:2006} emitted by the final \bh{},
extracted using least squares fitting.  As a cross-check, for some of
the models we also determine $a_\mathrm{f}/M_\mathrm{f}$ using an
approximate technique derived from the isolated horizon formalism~
\cite{2007arXiv0706.2541H,Ashtekar:2004cn}. Table~\ref{tbl:ED} gives
the energy $E_\mathrm{rad}$ and angular momentum $J_\mathrm{rad}$
radiated as well as the final mass $M_\mathrm{f}$ and spin
$a_\mathrm{f}$. Figure \ref{fig:massspin} gives the final mass
$M_\mathrm{f}$ and spin $a_\mathrm{f}$ as a function of $e$.  Notice
the agreement in $a_\mathrm{f}$ and $M_\mathrm{f}$ that the three
methods give within the estimated error bars.  The final mass and spin
also agree well in the circular case with the values obtained in
Ref.~\cite{2007PhRvD..76f4034B}.  The error bars for the
\qnm{}-derived quantities are dominated by the uncertainties in the
fitting procedure, which are estimated as the variations of the fit
parameters over a range of fitting windows.  The errors on the
radiation-balance quantities are dominated by the finite differencing
error of the simulations.  Due to excessive computational expense, we
have not run very high resolution versions of the eccentric
simulations, and so use the errors from the corresponding low
resolution circular orbit as a rough guide to the errors in the
eccentric cases.  

\begin{table}
  \begin{center}
\begin{ruledtabular}
\begin{tabular}{c|ccccccc}
$e$ & 
$\frac{E_\mathrm{rad}}{M_\mathrm{adm}}$ & $\frac{J_\mathrm{rad}}{M^2_\mathrm{adm}}$ & 
$\left.\frac{a_\mathrm{f}}{M_\mathrm{f}}\right\vert_\mathrm{rad}$ & 
$\left.\frac{a_\mathrm{f}}{M_\mathrm{f}}\right\vert_\mathrm{qnm}$ & 
$\left.\frac{a_\mathrm{f}}{M_\mathrm{f}}\right\vert_\mathrm{ih}$ &
$\left.\frac{M_\mathrm{f}}{M_\mathrm{adm}}\right\vert_\mathrm{rad}$ &
$\left.\frac{M_\mathrm{f}}{M_\mathrm{adm}}\right\vert_\mathrm{qnm}$ \\
\hline 
0.00 & 0.039 & 0.391 & 0.714 & 0.689 & -- & 0.961 & 0.964 \\ 
0.05 & 0.039 & 0.388 & 0.713 & 0.688 & -- & 0.961 & 0.963 \\ 
0.10 & 0.040 & 0.388 & 0.707 & 0.689 & -- & 0.960 & 0.963 \\ 
0.15 & 0.039 & 0.385 & 0.696 & 0.690 & -- & 0.961 & 0.964 \\ 
0.20 & 0.040 & 0.389 & 0.676 & 0.690 & -- & 0.960 & 0.963 \\ 
0.30 & 0.039 & 0.372 & 0.686 & 0.686 & 0.681 & 0.961 & 0.964 \\ 
0.40 & 0.040 & 0.279 & 0.716 & 0.698 & 0.693 & 0.960 & 0.962 \\ 
0.50 & 0.038 & 0.190 & 0.742 & 0.717 & 0.712 & 0.962 & 0.964 \\ 
0.60 & 0.022 & 0.108 & 0.713 & 0.707 & 0.702 & 0.978 & 0.980 \\ 
0.70 & 0.011 & 0.063 & 0.623 & 0.641 & 0.634 & 0.989 & 0.994 \\ 
0.80 & 0.004 & 0.033 & 0.484 & 0.515 & 0.502 & 0.996 & 1.002 \\ 
  \end{tabular}
  \end{ruledtabular}
  \end{center}
  \caption{\emph{Extracted quantities:} Energy  $E_\mathrm{rad}$ and angular momentum $J_\mathrm{rad}$ radiated;
    final spin parameter $a_\mathrm{f}$ and mass $M_\mathrm{f}$ 
    computed from $J_\mathrm{rad}$ and $E_\mathrm{rad}$ as well as from \qnm{} ringing.}
\label{tbl:ED}
\end{table}

Given an initial eccentricity, it is possible to choose a large enough
semimajor axis or orbital period for which the binary circularizes
before it arrives at the merger.  Our family of initial configurations
was designed to investigate, for a fixed initial orbital period, how
much initial eccentricity a binary is able to have and still enter the
merger with essentially vanishing eccentricity. Since we do not have a
good measure of eccentricity applicable prior to the merger, we focus
on the end state, namely $M_\mathrm{f}$ and $a_\mathrm{f}$ of the
final \bh{.}  We see from Fig.~\ref{fig:massspin} that
$a_\mathrm{f}/M_\mathrm{f} \approx 0.69$ for $e \lesssim 0.4$ and
$M_\mathrm{f}/M_\mathrm{adm} \approx 0.96$ for $e \lesssim 0.5$, both
values of $M_\mathrm{f}$ and $a_\mathrm{f}$ in agreement with the
circular result.  We note that the remaining orbits, the ones which do
not circularize, are all configurations which seem to plunge
immediately rather than entering an orbital phase.  We conclude that
for the systems we studied with approximately constant initial orbital
period, within our error bars, orbits with $e \lesssim 0.4$
essentially circularize before they merge, and orbits with $e \gtsim
0.5$ plunge.

We also observe for $e \gtsim 0.4$ that rather than $a_\mathrm{f}$
decreasing monotonically, a maximum spin parameter
$a_\mathrm{f}/M_\mathrm{f} \approx 0.72$ is obtained around $e = 0.5$.
Given the size of our uncertainties and that the maximum is found in
the three independent methods used to calculate the spin, we are
confident that this maximum is real for our family of initial data.
At about $e=0.6$, $a_\mathrm{f}$ starts decreasing monotonically.  We
are currently considering larger, but still computationally feasible,
initial separations to investigate if there is {\em any} bound orbital
(rather than plunge) configuration that does not circularize.

As $e \to 1$, corresponding to vanishing linear momenta (i.e.~a
head-on collision from rest), we find that $a_\mathrm{f}/M_\mathrm{f}
\to 0$, in line with the symmetry of the head-on collision, and $M_f
\sim M_\mathrm{adm}$, as expected, since NR simulations of a head-on
collision have shown that $M_\mathrm{f} \sim (1 - 0.001)
M_\mathrm{adm}$~\cite{Anninos:1993zj}.  Note that the ringdown result
for $e = 0.8$ gives $M_\mathrm{f} > M_\mathrm{adm}$ which is clearly
unphysical, but the error bars account for this.


{\em Conclusions:} We have carried out a series of eccentric orbit
simulations of \bbh{} systems in full nonlinear general relativity to
investigate the merger regime and final \bh{.}  The family of
simulations consisted of binaries with approximately constant initial
orbital period and varying initial eccentricity.  We found that for
initial $e \le 0.4$, the final \bh{} parameters are
$M_\mathrm{f}/M_\mathrm{adm} \approx 0.96$ and
$a_\mathrm{f}/M_\mathrm{f} \approx 0.69$, the same as in the circular
case. As a consequence of this, we also found that for $e \le 0.4$ the
binary begins to enter a \emph{universal} plunge at $t \sim
50\,M_\mathrm{f}$ before the amplitude of the gravitational radiation
reaches its peak.

While preparing the manuscript of this work, a study
by~\citet{Sperhake:2007gu} appeared with both similar and
complementary conclusions to those in our present work.  This work was
supported in part by NSF grants PHY-0354821, PHY-0653443, PHY-0244788,
PHY-0555436, PHY-0114375 (CGWP) and computer allocation TG-PHY060013N.
The authors thank M.~Ansorg, E.~Bentivegna, T.~Bode, A.~Knapp,
R.~Matzner and E.~Schnetter for contributions and helpful discussions,
and E.~Berti for the data tables used in the quasinormal
fitting.\vfill

\bibliography{references}

\end{document}

%% file: defs.tex
\def\ltsima{$\; \buildrel < \over \sim \;$}
\def\ltsim{\lower.5ex\hbox{\ltsima}}
\def\gtsima{$\; \buildrel > \over \sim \;$}
\def\gtsim{\lower.5ex\hbox{\gtsima}}

\def\newacronym#1#2#3{\gdef#1{#3 (#2)\gdef#1{#2}}}

\newacronym{\NSF}{NSF}{National Science Foundation}
\newacronym{\NASA}{NASA}{National Aeronautics and Space Administration}
\newacronym{\lisa}{LISA}{the Laser Interferometer Space Antenna}
\newacronym{\ligo}{LIGO}{Laser Interferometer Gravitational-wave Observatory} 
\newacronym{\Caltech}{Caltech}{California Institute of Technology}
\newacronym{\MIT}{MIT}{Massachusetts Institute of Technology}
\newacronym{\sph}{SPH}{smooth particle hydrodynamics}
\newacronym{\tsi}{TSI}{the Terascale Supernova Initiative}
\newacronym{\wmap}{WMAP}{the Wilkinson Microwave Anisotropy Probe}
\newacronym{\decigo}{DECIGO}{the Deci-Hertz Interferometric Gravitational-wave Observatory} 
\newacronym{\cmbr}{CMBR}{cosmic microwave background}
\newacronym{\ibbh}{IBBH}{intermediate binary black hole}
\newacronym{\bdj}{BDJ}{Brans-Dicke-Jordan}
\newacronym{\bbo}{BBO}{Big Bang Observer}
\newacronym{\decigo}{DECIGO}{Deci-Hertz Gravitational-Wave Observatory}

\def\MPR#1{{\it Moving Puncture Recipe}#1 (MPR#1)\gdef\MPR{MPR}}
\def\ahz#1{apparent horizon#1 (AH#1)\gdef\ahz{AH}}
\def\CM#1{center-of-mass#1 (CM#1)\gdef\CM{CM}}
\def\CLA#1{close-limit approximation#1 (CLA#1)\gdef\CLA{CLA}}
\def\pnw#1{post-Newtonian#1 (PN#1)\gdef\pnw{PN}}
\def\nr#1{numerical relativity#1 (NR#1)\gdef\nr{NR}}
\def\qnm#1{quasi-normal mode#1 (QNM#1)\gdef\qnm{QNM}}
\def\isco#1{innermost stable circular orbit#1 (ISCO#1)\gdef\isco{ISCO}}
\def\eos#1{equation of state#1 (EOS#1)\gdef\eos{EOS}}
\def\tov#1{Tolman-Oppenheimer-Volkoff#1 (TOV#1)\gdef\tov{TOV}}
\def\ns#1{neutron star#1 (NS#1)\gdef\ns{NS}}
\def\bbh#1{binary black hole#1 (BBH#1)\gdef\bbh{BBH}}
\def\bhns#1{black hole -- neutron star#1 (BHNS#1)\gdef\bhns{BHNS}}
\def\nsns#1{neutron star -- neutron star#1 (NSNS#1)\gdef\nsns{NSNS}}
\def\emri#1{extreme mass-ratio inspiral#1 (EMRI#1)\gdef\emri{EMRI}}
\def\emrb#1{extreme mass-ratio binaries#1 (EMRB#1)\gdef\emrb{EMRB}} 
\def\grb#1{gamma-ray burst#1 (GRB#1)\gdef\grb{GRB}}
\def\imbh#1{intermediate mass black hole#1 (IMBH#1)\gdef\imbh{IMBH}}
\def\smbh#1{supermassive black hole#1 (SMBH#1)\gdef\smbh{SMBH}}
\def\bh#1{black hole#1 (BH#1)\gdef\bh{BH}}
\def\ulx#1{ultra-luminous x-ray source#1 (ULX#1)\gdef\ulx{ULX}}
\def\nps#1{Newman-Penrose#1 (NP#1)\gdef\nps{NP}} 
\def\lmxbs{low-mass x-ray Binaries (LMXBs)\gdef\lmxbs{LMXBs}\gdef\lmxb{LMXB}} 
\def\lmxb{low-mass x-ray Binary (LMXB)\gdef\lmxbs{LMXBs}\gdef\lmxb{LMXB}}

\newcommand\apjl{\ref@jnl{ApJ}}%
\newcommand\mnras{\ref@jnl{MNRAS}}%

%% file: nrecc.bbl
\begin{thebibliography}{35}
\expandafter\ifx\csname natexlab\endcsname\relax\def\natexlab#1{#1}\fi
\expandafter\ifx\csname bibnamefont\endcsname\relax
  \def\bibnamefont#1{#1}\fi
\expandafter\ifx\csname bibfnamefont\endcsname\relax
  \def\bibfnamefont#1{#1}\fi
\expandafter\ifx\csname citenamefont\endcsname\relax
  \def\citenamefont#1{#1}\fi
\expandafter\ifx\csname url\endcsname\relax
  \def\url#1{\texttt{#1}}\fi
\expandafter\ifx\csname urlprefix\endcsname\relax\def\urlprefix{URL }\fi
\providecommand{\bibinfo}[2]{#2}
\providecommand{\eprint}[2][]{\url{#2}}

\bibitem[{\citenamefont{Baker et~al.}(2006)\citenamefont{Baker, Centrella,
  Choi, Koppitz, and van Meter}}]{Baker:2006yw}
\bibinfo{author}{\bibfnamefont{J.~G.} \bibnamefont{Baker}},
  \bibinfo{author}{\bibfnamefont{J.}~\bibnamefont{Centrella}},
  \bibinfo{author}{\bibfnamefont{D.-I.} \bibnamefont{Choi}},
  \bibinfo{author}{\bibfnamefont{M.}~\bibnamefont{Koppitz}}, \bibnamefont{and}
  \bibinfo{author}{\bibfnamefont{J.}~\bibnamefont{van Meter}},
  \bibinfo{journal}{Phys. Rev.} \textbf{\bibinfo{volume}{D73}},
  \bibinfo{pages}{104002} (\bibinfo{year}{2006}).

\bibitem[{\citenamefont{Hannam et~al.}(2007)\citenamefont{Hannam, Husa,
  Sperhake, Br{\"u}gmann, and Gonzalez}}]{Hannam:2007ik}
\bibinfo{author}{\bibfnamefont{M.}~\bibnamefont{Hannam}},
  \bibinfo{author}{\bibfnamefont{S.}~\bibnamefont{Husa}},
  \bibinfo{author}{\bibfnamefont{U.}~\bibnamefont{Sperhake}},
  \bibinfo{author}{\bibfnamefont{B.}~\bibnamefont{Br{\"u}gmann}},
  \bibnamefont{and} \bibinfo{author}{\bibfnamefont{J.~A.}
  \bibnamefont{Gonzalez}}, \bibinfo{journal}{preprint (arXiv:0706.1305)}
  (\bibinfo{year}{2007}).

\bibitem[{\citenamefont{Boyle et~al.}(2007)\citenamefont{Boyle, Brown, Kidder,
  Mroue, Pfeiffer, Scheel, Cook, and Teukolsky}}]{boyle-2007}
\bibinfo{author}{\bibfnamefont{M.}~\bibnamefont{Boyle}},
  \bibinfo{author}{\bibfnamefont{D.~A.} \bibnamefont{Brown}},
  \bibinfo{author}{\bibfnamefont{L.~E.} \bibnamefont{Kidder}},
  \bibinfo{author}{\bibfnamefont{A.~H.} \bibnamefont{Mroue}},
  \bibinfo{author}{\bibfnamefont{H.~P.} \bibnamefont{Pfeiffer}},
  \bibinfo{author}{\bibfnamefont{M.~A.} \bibnamefont{Scheel}},
  \bibinfo{author}{\bibfnamefont{G.~B.} \bibnamefont{Cook}}, \bibnamefont{and}
  \bibinfo{author}{\bibfnamefont{S.~A.} \bibnamefont{Teukolsky}},
  \bibinfo{journal}{preprint (arXiv.org:0710.0158)}  (\bibinfo{year}{2007}).

\bibitem[{\citenamefont{{Campanelli} et~al.}(2007)\citenamefont{{Campanelli},
  {Lousto}, {Zlochower}, and {Merritt}}}]{2007gr.qc.....1164C}
\bibinfo{author}{\bibfnamefont{M.}~\bibnamefont{{Campanelli}}},
  \bibinfo{author}{\bibfnamefont{C.~O.} \bibnamefont{{Lousto}}},
  \bibinfo{author}{\bibfnamefont{Y.}~\bibnamefont{{Zlochower}}},
  \bibnamefont{and}
  \bibinfo{author}{\bibfnamefont{D.}~\bibnamefont{{Merritt}}},
  \bibinfo{journal}{Ap. J. Lett.} \textbf{\bibinfo{volume}{659}},
  \bibinfo{pages}{L5} (\bibinfo{year}{2007}).

\bibitem[{\citenamefont{Husa et~al.}(2007{\natexlab{a}})\citenamefont{Husa,
  Gonzalez, Hannam, Br{\"u}gmann, and Sperhake}}]{husa-2007}
\bibinfo{author}{\bibfnamefont{S.}~\bibnamefont{Husa}},
  \bibinfo{author}{\bibfnamefont{J.~A.} \bibnamefont{Gonzalez}},
  \bibinfo{author}{\bibfnamefont{M.}~\bibnamefont{Hannam}},
  \bibinfo{author}{\bibfnamefont{B.}~\bibnamefont{Br{\"u}gmann}},
  \bibnamefont{and} \bibinfo{author}{\bibfnamefont{U.}~\bibnamefont{Sperhake}},
  \bibinfo{journal}{preprint (arXiv.org:0706.0740)}
  (\bibinfo{year}{2007}{\natexlab{a}}).

\bibitem[{\citenamefont{{Gonzalez} et~al.}(2007)\citenamefont{{Gonzalez},
  {Sperhake}, {Br{\"u}gmann}, {Hannam}, and {Husa}}}]{Gonzalez:2006md}
\bibinfo{author}{\bibfnamefont{J.~A.} \bibnamefont{{Gonzalez}}},
  \bibinfo{author}{\bibfnamefont{U.}~\bibnamefont{{Sperhake}}},
  \bibinfo{author}{\bibfnamefont{B.}~\bibnamefont{{Br{\"u}gmann}}},
  \bibinfo{author}{\bibfnamefont{M.}~\bibnamefont{{Hannam}}}, \bibnamefont{and}
  \bibinfo{author}{\bibfnamefont{S.}~\bibnamefont{{Husa}}},
  \bibinfo{journal}{Phys. Rev. Lett.} \textbf{\bibinfo{volume}{98}},
  \bibinfo{pages}{091101} (\bibinfo{year}{2007}).

\bibitem[{\citenamefont{{Baker} et~al.}(2006)\citenamefont{{Baker},
  {Centrella}, {Choi}, {Koppitz}, {van Meter}, and {Miller}}}]{Baker:2006vn}
\bibinfo{author}{\bibfnamefont{J.~G.} \bibnamefont{{Baker}}},
  \bibinfo{author}{\bibfnamefont{J.}~\bibnamefont{{Centrella}}},
  \bibinfo{author}{\bibfnamefont{D.-I.} \bibnamefont{{Choi}}},
  \bibinfo{author}{\bibfnamefont{M.}~\bibnamefont{{Koppitz}}},
  \bibinfo{author}{\bibfnamefont{J.~R.} \bibnamefont{{van Meter}}},
  \bibnamefont{and} \bibinfo{author}{\bibfnamefont{M.~C.}
  \bibnamefont{{Miller}}}, \bibinfo{journal}{Ap. J. Lett.}
  \textbf{\bibinfo{volume}{653}}, \bibinfo{pages}{L93} (\bibinfo{year}{2006}).

\bibitem[{\citenamefont{Gonzalez et~al.}(2007)\citenamefont{Gonzalez, Hannam,
  Sperhake, Br{\"u}gmann, and Husa}}]{Gonzalez:2007hi}
\bibinfo{author}{\bibfnamefont{J.~A.} \bibnamefont{Gonzalez}},
  \bibinfo{author}{\bibfnamefont{M.~D.} \bibnamefont{Hannam}},
  \bibinfo{author}{\bibfnamefont{U.}~\bibnamefont{Sperhake}},
  \bibinfo{author}{\bibfnamefont{B.}~\bibnamefont{Br{\"u}gmann}},
  \bibnamefont{and} \bibinfo{author}{\bibfnamefont{S.}~\bibnamefont{Husa}},
  \bibinfo{journal}{Phys. Rev. Lett.} \textbf{\bibinfo{volume}{98}},
  \bibinfo{pages}{231101} (\bibinfo{year}{2007}), \eprint{gr-qc/0702052}.

\bibitem[{\citenamefont{{Herrmann}
  et~al.}(2007{\natexlab{a}})\citenamefont{{Herrmann}, {Hinder}, {Shoemaker},
  {Laguna}, and {Matzner}}}]{2007ApJ...661..430H}
\bibinfo{author}{\bibfnamefont{F.}~\bibnamefont{{Herrmann}}},
  \bibinfo{author}{\bibfnamefont{I.}~\bibnamefont{{Hinder}}},
  \bibinfo{author}{\bibfnamefont{D.}~\bibnamefont{{Shoemaker}}},
  \bibinfo{author}{\bibfnamefont{P.}~\bibnamefont{{Laguna}}}, \bibnamefont{and}
  \bibinfo{author}{\bibfnamefont{R.~A.} \bibnamefont{{Matzner}}},
  \bibinfo{journal}{Aptrophys J.} \textbf{\bibinfo{volume}{661}},
  \bibinfo{pages}{430} (\bibinfo{year}{2007}{\natexlab{a}}).

\bibitem[{\citenamefont{{Koppitz} et~al.}(2007)\citenamefont{{Koppitz},
  {Pollney}, {Reisswig}, {Rezzolla}, {Thornburg}, {Diener}, and
  {Schnetter}}}]{2007PhRvL..99d1102K}
\bibinfo{author}{\bibfnamefont{M.}~\bibnamefont{{Koppitz}}},
  \bibinfo{author}{\bibfnamefont{D.}~\bibnamefont{{Pollney}}},
  \bibinfo{author}{\bibfnamefont{C.}~\bibnamefont{{Reisswig}}},
  \bibinfo{author}{\bibfnamefont{L.}~\bibnamefont{{Rezzolla}}},
  \bibinfo{author}{\bibfnamefont{J.}~\bibnamefont{{Thornburg}}},
  \bibinfo{author}{\bibfnamefont{P.}~\bibnamefont{{Diener}}}, \bibnamefont{and}
  \bibinfo{author}{\bibfnamefont{E.}~\bibnamefont{{Schnetter}}},
  \bibinfo{journal}{\prl} \textbf{\bibinfo{volume}{99}},
  \bibinfo{pages}{041102} (\bibinfo{year}{2007}), \eprint{arXiv:gr-qc/0701163}.

\bibitem[{\citenamefont{Campanelli et~al.}(2007)\citenamefont{Campanelli,
  Lousto, Zlochower, and Merritt}}]{Campanelli:2007cga}
\bibinfo{author}{\bibfnamefont{M.}~\bibnamefont{Campanelli}},
  \bibinfo{author}{\bibfnamefont{C.~O.} \bibnamefont{Lousto}},
  \bibinfo{author}{\bibfnamefont{Y.}~\bibnamefont{Zlochower}},
  \bibnamefont{and} \bibinfo{author}{\bibfnamefont{D.}~\bibnamefont{Merritt}},
  \bibinfo{journal}{Phys. Rev. Lett.} \textbf{\bibinfo{volume}{98}},
  \bibinfo{pages}{231102} (\bibinfo{year}{2007}), \eprint{gr-qc/0702133}.

\bibitem[{\citenamefont{Campanelli
  et~al.}(2006{\natexlab{a}})\citenamefont{Campanelli, Lousto, and
  Zlochower}}]{Campanelli:2006uy}
\bibinfo{author}{\bibfnamefont{M.}~\bibnamefont{Campanelli}},
  \bibinfo{author}{\bibfnamefont{C.~O.} \bibnamefont{Lousto}},
  \bibnamefont{and}
  \bibinfo{author}{\bibfnamefont{Y.}~\bibnamefont{Zlochower}},
  \bibinfo{journal}{Phys. Rev. D} \textbf{\bibinfo{volume}{74}},
  \bibinfo{pages}{041501} (\bibinfo{year}{2006}{\natexlab{a}}).

\bibitem[{\citenamefont{{Herrmann}
  et~al.}(2007{\natexlab{b}})\citenamefont{{Herrmann}, {Hinder}, {Shoemaker},
  {Laguna}, and {Matzner}}}]{2007arXiv0706.2541H}
\bibinfo{author}{\bibfnamefont{F.}~\bibnamefont{{Herrmann}}},
  \bibinfo{author}{\bibfnamefont{I.}~\bibnamefont{{Hinder}}},
  \bibinfo{author}{\bibfnamefont{D.~M.} \bibnamefont{{Shoemaker}}},
  \bibinfo{author}{\bibfnamefont{P.}~\bibnamefont{{Laguna}}}, \bibnamefont{and}
  \bibinfo{author}{\bibfnamefont{R.~A.} \bibnamefont{{Matzner}}},
  \bibinfo{journal}{preprint (arXiv:0706.2541)}
  (\bibinfo{year}{2007}{\natexlab{b}}).

\bibitem[{\citenamefont{Pretorius}(2006)}]{pretorius-2006-23}
\bibinfo{author}{\bibfnamefont{F.}~\bibnamefont{Pretorius}},
  \bibinfo{journal}{Class. Quant. Grav.} \textbf{\bibinfo{volume}{23}},
  \bibinfo{pages}{S529} (\bibinfo{year}{2006}).

\bibitem[{\citenamefont{Pretorius and Khurana}(2007)}]{pretorius-2007}
\bibinfo{author}{\bibfnamefont{F.}~\bibnamefont{Pretorius}} \bibnamefont{and}
  \bibinfo{author}{\bibfnamefont{D.}~\bibnamefont{Khurana}},
  \bibinfo{journal}{preprint (gr-qc/0702084)}  (\bibinfo{year}{2007}).

\bibitem[{\citenamefont{{Pollney} et~al.}(2007)\citenamefont{{Pollney},
  {Reisswig}, {Rezzolla}, {Szilagyi}, {Ansorg}, {Deris}, {Diener}, {Dorband},
  {Koppitz}, {Nagar} et~al.}}]{2007arXiv0707.2559P}
\bibinfo{author}{\bibfnamefont{D.}~\bibnamefont{{Pollney}}},
  \bibinfo{author}{\bibfnamefont{C.}~\bibnamefont{{Reisswig}}},
  \bibinfo{author}{\bibfnamefont{L.}~\bibnamefont{{Rezzolla}}},
  \bibinfo{author}{\bibfnamefont{B.}~\bibnamefont{{Szilagyi}}},
  \bibinfo{author}{\bibfnamefont{M.}~\bibnamefont{{Ansorg}}},
  \bibinfo{author}{\bibfnamefont{B.}~\bibnamefont{{Deris}}},
  \bibinfo{author}{\bibfnamefont{P.}~\bibnamefont{{Diener}}},
  \bibinfo{author}{\bibfnamefont{E.~N.} \bibnamefont{{Dorband}}},
  \bibinfo{author}{\bibfnamefont{M.}~\bibnamefont{{Koppitz}}},
  \bibinfo{author}{\bibfnamefont{A.}~\bibnamefont{{Nagar}}},
  \bibnamefont{et~al.}, \bibinfo{journal}{ArXiv e-prints}
  \textbf{\bibinfo{volume}{707}} (\bibinfo{year}{2007}), \eprint{0707.2559}.

\bibitem[{\citenamefont{{Rezzolla}
  et~al.}(2007{\natexlab{a}})\citenamefont{{Rezzolla}, {Dorband}, {Reisswig},
  {Diener}, {Pollney}, {Schnetter}, and {Szilagyi}}}]{2007arXiv0708.3999R}
\bibinfo{author}{\bibfnamefont{L.}~\bibnamefont{{Rezzolla}}},
  \bibinfo{author}{\bibfnamefont{E.~N.} \bibnamefont{{Dorband}}},
  \bibinfo{author}{\bibfnamefont{C.}~\bibnamefont{{Reisswig}}},
  \bibinfo{author}{\bibfnamefont{P.}~\bibnamefont{{Diener}}},
  \bibinfo{author}{\bibfnamefont{D.}~\bibnamefont{{Pollney}}},
  \bibinfo{author}{\bibfnamefont{E.}~\bibnamefont{{Schnetter}}},
  \bibnamefont{and}
  \bibinfo{author}{\bibfnamefont{B.}~\bibnamefont{{Szilagyi}}},
  \bibinfo{journal}{ArXiv e-prints} \textbf{\bibinfo{volume}{708}}
  (\bibinfo{year}{2007}{\natexlab{a}}), \eprint{0708.3999}.

\bibitem[{\citenamefont{{Rezzolla}
  et~al.}(2007{\natexlab{b}})\citenamefont{{Rezzolla}, {Diener}, {Dorband},
  {Pollney}, {Reisswig}, {Schnetter}, and {Seiler}}}]{2007arXiv0710.3345R}
\bibinfo{author}{\bibfnamefont{L.}~\bibnamefont{{Rezzolla}}},
  \bibinfo{author}{\bibfnamefont{P.}~\bibnamefont{{Diener}}},
  \bibinfo{author}{\bibfnamefont{E.~N.} \bibnamefont{{Dorband}}},
  \bibinfo{author}{\bibfnamefont{D.}~\bibnamefont{{Pollney}}},
  \bibinfo{author}{\bibfnamefont{C.}~\bibnamefont{{Reisswig}}},
  \bibinfo{author}{\bibfnamefont{E.}~\bibnamefont{{Schnetter}}},
  \bibnamefont{and} \bibinfo{author}{\bibfnamefont{J.}~\bibnamefont{{Seiler}}},
  \bibinfo{journal}{ArXiv e-prints} \textbf{\bibinfo{volume}{710}}
  (\bibinfo{year}{2007}{\natexlab{b}}), \eprint{0710.3345}.

\bibitem[{\citenamefont{Br{\"u}gmann et~al.}(2004)\citenamefont{Br{\"u}gmann,
  Tichy, and Jansen}}]{Bruegmann:2003aw}
\bibinfo{author}{\bibfnamefont{B.}~\bibnamefont{Br{\"u}gmann}},
  \bibinfo{author}{\bibfnamefont{W.}~\bibnamefont{Tichy}}, \bibnamefont{and}
  \bibinfo{author}{\bibfnamefont{N.}~\bibnamefont{Jansen}},
  \bibinfo{journal}{Phys. Rev. Lett.} \textbf{\bibinfo{volume}{92}},
  \bibinfo{pages}{211101} (\bibinfo{year}{2004}).

\bibitem[{\citenamefont{{Pretorius}}(2005)}]{2005PhRvL..95l1101P}
\bibinfo{author}{\bibfnamefont{F.}~\bibnamefont{{Pretorius}}},
  \bibinfo{journal}{Phys. Rev. Lett.} \textbf{\bibinfo{volume}{95}},
  \bibinfo{pages}{121101} (\bibinfo{year}{2005}).

\bibitem[{\citenamefont{Baker et~al.}(2006)\citenamefont{Baker, Centrella,
  Choi, Koppitz, and van Meter}}]{Baker:2005vv}
\bibinfo{author}{\bibfnamefont{J.~G.} \bibnamefont{Baker}},
  \bibinfo{author}{\bibfnamefont{J.}~\bibnamefont{Centrella}},
  \bibinfo{author}{\bibfnamefont{D.-I.} \bibnamefont{Choi}},
  \bibinfo{author}{\bibfnamefont{M.}~\bibnamefont{Koppitz}}, \bibnamefont{and}
  \bibinfo{author}{\bibfnamefont{J.}~\bibnamefont{van Meter}},
  \bibinfo{journal}{Phys. Rev. Lett.} \textbf{\bibinfo{volume}{96}},
  \bibinfo{pages}{111102} (\bibinfo{year}{2006}).

\bibitem[{\citenamefont{Campanelli
  et~al.}(2006{\natexlab{b}})\citenamefont{Campanelli, Lousto, Marronetti, and
  Zlochower}}]{Campanelli:2005dd}
\bibinfo{author}{\bibfnamefont{M.}~\bibnamefont{Campanelli}},
  \bibinfo{author}{\bibfnamefont{C.~O.} \bibnamefont{Lousto}},
  \bibinfo{author}{\bibfnamefont{P.}~\bibnamefont{Marronetti}},
  \bibnamefont{and}
  \bibinfo{author}{\bibfnamefont{Y.}~\bibnamefont{Zlochower}},
  \bibinfo{journal}{Phys. Rev. Lett.} \textbf{\bibinfo{volume}{96}},
  \bibinfo{pages}{111101} (\bibinfo{year}{2006}{\natexlab{b}}).

\bibitem[{\citenamefont{Peters}(1964)}]{Peters:1964}
\bibinfo{author}{\bibfnamefont{P.~C.} \bibnamefont{Peters}},
  \bibinfo{journal}{Phys. Rev.} \textbf{\bibinfo{volume}{136}},
  \bibinfo{pages}{B1224} (\bibinfo{year}{1964}).

\bibitem[{\citenamefont{{Miller} and {Hamilton}}(2002)}]{2002ApJ...576..894M}
\bibinfo{author}{\bibfnamefont{M.~C.} \bibnamefont{{Miller}}} \bibnamefont{and}
  \bibinfo{author}{\bibfnamefont{D.~P.} \bibnamefont{{Hamilton}}},
  \bibinfo{journal}{\apj} \textbf{\bibinfo{volume}{576}}, \bibinfo{pages}{894}
  (\bibinfo{year}{2002}).

\bibitem[{\citenamefont{{Wen}}(2003)}]{2003ApJ...598..419W}
\bibinfo{author}{\bibfnamefont{L.}~\bibnamefont{{Wen}}},
  \bibinfo{journal}{\apj} \textbf{\bibinfo{volume}{598}}, \bibinfo{pages}{419}
  (\bibinfo{year}{2003}).

\bibitem[{\citenamefont{{Armitage} and
  {Natarajan}}(2005)}]{2005ApJ...634..921A}
\bibinfo{author}{\bibfnamefont{P.~J.} \bibnamefont{{Armitage}}}
  \bibnamefont{and}
  \bibinfo{author}{\bibfnamefont{P.}~\bibnamefont{{Natarajan}}},
  \bibinfo{journal}{\apj} \textbf{\bibinfo{volume}{634}}, \bibinfo{pages}{921}
  (\bibinfo{year}{2005}).

\bibitem[{\citenamefont{Anninos et~al.}(1995)\citenamefont{Anninos, Price,
  Pullin, Seidel, and Suen}}]{Anninos:1995vf}
\bibinfo{author}{\bibfnamefont{P.}~\bibnamefont{Anninos}},
  \bibinfo{author}{\bibfnamefont{R.~H.} \bibnamefont{Price}},
  \bibinfo{author}{\bibfnamefont{J.}~\bibnamefont{Pullin}},
  \bibinfo{author}{\bibfnamefont{E.}~\bibnamefont{Seidel}}, \bibnamefont{and}
  \bibinfo{author}{\bibfnamefont{W.-M.} \bibnamefont{Suen}},
  \bibinfo{journal}{Phys. Rev. D} \textbf{\bibinfo{volume}{52}},
  \bibinfo{pages}{4462} (\bibinfo{year}{1995}).

\bibitem[{\citenamefont{Husa et~al.}(2007{\natexlab{b}})\citenamefont{Husa,
  Hannam, Gonzalez, Sperhake, and Br{\"u}gmann}}]{Husa:2007rh}
\bibinfo{author}{\bibfnamefont{S.}~\bibnamefont{Husa}},
  \bibinfo{author}{\bibfnamefont{M.}~\bibnamefont{Hannam}},
  \bibinfo{author}{\bibfnamefont{J.~A.} \bibnamefont{Gonzalez}},
  \bibinfo{author}{\bibfnamefont{U.}~\bibnamefont{Sperhake}}, \bibnamefont{and}
  \bibinfo{author}{\bibfnamefont{B.}~\bibnamefont{Br{\"u}gmann}},
  \bibinfo{journal}{preprint (arXiv:0706.0904)}
  (\bibinfo{year}{2007}{\natexlab{b}}).

\bibitem[{\citenamefont{{K{\"o}nigsd{\"o}rffer} and
  {Gopakumar}}(2006)}]{2006PhRvD..73l4012K}
\bibinfo{author}{\bibfnamefont{C.}~\bibnamefont{{K{\"o}nigsd{\"o}rffer}}}
  \bibnamefont{and}
  \bibinfo{author}{\bibfnamefont{A.}~\bibnamefont{{Gopakumar}}},
  \bibinfo{journal}{\prd} \textbf{\bibinfo{volume}{73}},
  \bibinfo{pages}{124012} (\bibinfo{year}{2006}).

\bibitem[{\citenamefont{Vaishnav et~al.}(2007)\citenamefont{Vaishnav, Hinder,
  Herrmann, and Shoemaker}}]{vaishnav-2007}
\bibinfo{author}{\bibfnamefont{B.}~\bibnamefont{Vaishnav}},
  \bibinfo{author}{\bibfnamefont{I.}~\bibnamefont{Hinder}},
  \bibinfo{author}{\bibfnamefont{F.}~\bibnamefont{Herrmann}}, \bibnamefont{and}
  \bibinfo{author}{\bibfnamefont{D.}~\bibnamefont{Shoemaker}},
  \bibinfo{journal}{Phys. Rev. D} \textbf{\bibinfo{volume}{76}},
  \bibinfo{pages}{084020} (\bibinfo{year}{2007}).

\bibitem[{\citenamefont{Berti et~al.}(2006)\citenamefont{Berti, Cardoso, and
  Will}}]{BertiLISA:2006}
\bibinfo{author}{\bibfnamefont{E.}~\bibnamefont{Berti}},
  \bibinfo{author}{\bibfnamefont{V.}~\bibnamefont{Cardoso}}, \bibnamefont{and}
  \bibinfo{author}{\bibfnamefont{C.~M.} \bibnamefont{Will}},
  \bibinfo{journal}{Phys. Rev. D} \textbf{\bibinfo{volume}{73}},
  \bibinfo{pages}{064030} (\bibinfo{year}{2006}).

\bibitem[{\citenamefont{Ashtekar and Krishnan}(2004)}]{Ashtekar:2004cn}
\bibinfo{author}{\bibfnamefont{A.}~\bibnamefont{Ashtekar}} \bibnamefont{and}
  \bibinfo{author}{\bibfnamefont{B.}~\bibnamefont{Krishnan}},
  \bibinfo{journal}{Living Rev. Rel.} \textbf{\bibinfo{volume}{7}},
  \bibinfo{pages}{10} (\bibinfo{year}{2004}), \eprint{gr-qc/0407042}.

\bibitem[{\citenamefont{{Berti} et~al.}(2007)\citenamefont{{Berti}, {Cardoso},
  {Gonzalez}, {Sperhake}, {Hannam}, {Husa}, and
  {Br{\"u}gmann}}}]{2007PhRvD..76f4034B}
\bibinfo{author}{\bibfnamefont{E.}~\bibnamefont{{Berti}}},
  \bibinfo{author}{\bibfnamefont{V.}~\bibnamefont{{Cardoso}}},
  \bibinfo{author}{\bibfnamefont{J.~A.} \bibnamefont{{Gonzalez}}},
  \bibinfo{author}{\bibfnamefont{U.}~\bibnamefont{{Sperhake}}},
  \bibinfo{author}{\bibfnamefont{M.}~\bibnamefont{{Hannam}}},
  \bibinfo{author}{\bibfnamefont{S.}~\bibnamefont{{Husa}}}, \bibnamefont{and}
  \bibinfo{author}{\bibfnamefont{B.}~\bibnamefont{{Br{\"u}gmann}}},
  \bibinfo{journal}{\prd} \textbf{\bibinfo{volume}{76}},
  \bibinfo{pages}{064034} (\bibinfo{year}{2007}), \eprint{arXiv:gr-qc/0703053}.

\bibitem[{\citenamefont{Anninos et~al.}(1993)\citenamefont{Anninos, Hobill,
  Seidel, Smarr, and Suen}}]{Anninos:1993zj}
\bibinfo{author}{\bibfnamefont{P.}~\bibnamefont{Anninos}},
  \bibinfo{author}{\bibfnamefont{D.}~\bibnamefont{Hobill}},
  \bibinfo{author}{\bibfnamefont{E.}~\bibnamefont{Seidel}},
  \bibinfo{author}{\bibfnamefont{L.}~\bibnamefont{Smarr}}, \bibnamefont{and}
  \bibinfo{author}{\bibfnamefont{W.-M.} \bibnamefont{Suen}},
  \bibinfo{journal}{Phys. Rev. Lett.} \textbf{\bibinfo{volume}{71}},
  \bibinfo{pages}{2851} (\bibinfo{year}{1993}).

\bibitem[{\citenamefont{Sperhake et~al.}(2007)\citenamefont{Sperhake, Berti,
  Cardoso, Gonzalez, Br{\"u}gmann, and Ansorg}}]{Sperhake:2007gu}
\bibinfo{author}{\bibfnamefont{U.}~\bibnamefont{Sperhake}},
  \bibinfo{author}{\bibfnamefont{E.}~\bibnamefont{Berti}},
  \bibinfo{author}{\bibfnamefont{V.}~\bibnamefont{Cardoso}},
  \bibinfo{author}{\bibfnamefont{J.}~\bibnamefont{Gonzalez}},
  \bibinfo{author}{\bibfnamefont{B.}~\bibnamefont{Br{\"u}gmann}},
  \bibnamefont{and} \bibinfo{author}{\bibfnamefont{M.}~\bibnamefont{Ansorg}},
  \bibinfo{journal}{preprint (arXiv:0710.3823)}  (\bibinfo{year}{2007}).

\end{thebibliography}
